\documentclass[sigconf]{acmart}

\usepackage{booktabs} 

\setcopyright{none}

\acmYear{2023}
\setcopyright{sandia}
\acmConference[NICE 2023]{Neuro-Inspired Computational Elements Conference}{April 11-14, 2023}{San Antonio, TX, USA}
\acmBooktitle{Neuro-Inspired Computational Elements Conference (NICE 2023), as poster, April 11-14, 2023, San Antonio, TX, USA}
\acmDOI{}
\acmISBN{}

\begin{document}
\title[]{Distributed Compressed Sparse Row Format for Spiking Neural Network Simulation, Serialization, and Interoperability}

\author{Felix Wang}
\orcid{0000-0002-4288-2238}
\affiliation{%
  \institution{Sandia National Laboratories}
  \city{Albuquerque}
  \state{New Mexico}
  \country{USA}
  \postcode{87123}
}
\email{felwang@sandia.gov}

\renewcommand{\shortauthors}{Felix Wang}

\begin{abstract}
With the increasing development of neuromorphic platforms and their related software tools as well as the increasing scale of spiking neural network (SNN) models, there is a pressure for interoperable and scalable representations of network state. In response to this, we discuss a parallel extension of a widely used format for efficiently representing sparse matrices, the compressed sparse row (CSR), in the context of supporting the simulation and serialization of large-scale SNNs. Sparse matrices for graph adjacency structure provide a natural fit for describing the connectivity of an SNN, and prior work in the area of parallel graph partitioning has developed the distributed CSR (dCSR) format for storing and ingesting large graphs. We contend that organizing additional network information, such as neuron and synapse state, in alignment with its adjacency as dCSR provides a straightforward partition-based distribution of network state. For large-scale simulations, this means each parallel process is only responsible for its own partition of state, which becomes especially useful when the size of an SNN exceeds the memory resources of a single compute node. For potentially long-running simulations, this also enables network serialization to and from disk (e.g. for checkpoint/restart fault-tolerant computing) to be performed largely independently between parallel processes. We also provide a potential implementation, and put it forward for adoption within the neural computing community.
\end{abstract}


\maketitle

\section{Introduction}
The field of neural computing has witnessed significant developments and expansions in the software frameworks, network simulators, hardware platforms, and engineering tools available to the community~\cite{Schuman:2017,Dai:2020b,Stimberg:2019,Aimone:2019,Lava:2022,Rothganger:2014,Davison:2009,Brette:2007}. Especially in the software space, there has also been a continual push toward larger scale experiments in line with the advancements in high performance computing (HPC) and use of accelerators such as graphics processing units (GPU)~\cite{Golosio:2021,Knight:2021,Norse:2021,Wang:2015}. However, alongside this expansion of frameworks and tools is the question of their compatibility and interoperability within a shared ecosystem. For example, a common thread of discussion is how to adequately compare and benchmark between different simulators and target neuromorphic platforms~\cite{Vineyard:2019,Kulkarni:2021,Davies:2019}.

Recent efforts such as Fugu have attempted to resolve the interoperability between a spiking neural algorithm and different neurormophic backends by employing only the most widely supported neuron and synapse models~\cite{Aimone:2019}. In the adjacent machine learning (ML) community, there are also efforts such as the open neural network exchange (ONNX), which provides an open format built to represent ML models (e.g. between PyTorch and TensorFlow)~\cite{Bai:2019}. Perhaps most closely related to this work is the Scalable Open Network Architecture TemplAte (SONATA) data format for large-scale network models~\cite{Dai:2020}. Although it leverages common file formats (e.g. csv, hdf5, json), we noted some concerns due to its use of hierarchical, population-based grouping (for both neurons and synapses), especially with regards to data locality.

In this paper, we propose the use of a relatively straightforward data format for interoperability and sharing of SNN models between simulators and neuromorphic hardware platforms. We base this on the widely used compressed sparse row (CSR) format for efficiently representing sparse matrices, and extend it to accommodate parallel partitions of network state. We first provide a high-level description of this format in Section \ref{sec:overview}, a pointer to an implementation in Section \ref{sec:specification}, and finally a discussion in Section \ref{sec:discussion}.

\section{Overview}\label{sec:overview}

Compressed sparse row is a widely used format to efficiently represent sparse matrices~\cite{Saad:2011}. Although there are several implementations, the central idea is to store the non-zero values of a matrix corresponding to indexical arrays over the rows and columns. For an $(n\times n)$ matrix with $m$ non-zero entries, the row array is of size $n+1$, where each entry provides a prefix or cumulative sum over the number of non-zero column indices for that row, and the last entry in the row array is $m$. The column array is of size $m$ and contains the indices of the non-zeros entries for a given row-column pair. The corresponding value array is also of size $m$ and simply concatenates the non-zero values as read off in row-major order.

As an extension to this, the distributed CSR (dCSR) format was most notably introduced for storing and ingesting large graphs in prior work in parallel graph partitioning algorithms~\cite{Karypis:1998}. This format provides an additional indexical array of size $k+1$ which supplies the prefix or cumulative sum over the number of non-zero indices for a given $k$-way partitioning of the rows, where $n_1 + n_2 + \dots + n_k = n$. Furthermore, the original CSR value and column arrays are split into multiple arrays along these partitions and $m_1 + m_2 + \dots + m_k = m$.

To compare, for an SNN, a natural representation of the connectivity and overall network structure is as a directed graph~\cite{West:2001}. Vertices $V(G)=\{v_1, v_2, \dots, v_n\}$ correspond to the spiking neurons, and edges $E(G)=\{e_1, e_2, \dots, e_m\}$ correspond to the synaptic connections. The direction of an edge $e_l:v_i \to v_j$, for $i,j\in1,\dots,n$ and $l\in1,\dots,m$ corresponds to the propagation of an event (e.g. a spike) from the presynaptic neuron $v_i$ to the postsynaptic neuron $v_j$. We may further partition this graph by splitting up the vertices such that $|V_1| + |V_2| + \dots + |V_k| = |V|$, known as a $k$-way partition. Here, edges may be assigned to a given partition based on their source vertex or their target vertex. With respect to the communication and computational patterns for SNNs, typically with synaptic weights applying current on their target neuron, colocating a directed edge with its target vertex is more sensible.

We can see that there is essentially an isomorphism between the rows and columns of a sparse matrix to the vertices and edges of an SNN graph. The main difference is that for an SNN, the vertices and edges typically carry far more additional information (e.g. connection delays, neural and synaptic states) than may be afforded by the standard, single non-zero entry in the value array. To address this, we suggest the extension to simply allow for tuples of values to be associated with the column array (as well as tuples of values to be associated with the row array). Because the amount of necessary unique state for any given vertex or edge will depend on its specific model dynamics, we may also introduce an additional model dictionary to provide tuple sizes.

\section{Serialization Format}\label{sec:specification}
Transitioning from this high-level description to a more concrete implementation is actually fairly straightforward. In fact, we may simply extend the dCSR format such as used by ParMETIS~\cite{Parmetis:2013}. This has the added advantage that we may directly interoperate with existing graph partitioning tools. Although generally less memory efficient on-disk than in simulation, we also opt to serialize to plain-text files for portability.

Instead of serializing a graph adjacency as two contiguous arrays of row offsets and column indices, respectively, one of the shortcuts implemented by ParMETIS is to implicitly index the row by its line number in a given \texttt{.adjcy.k} (adjacency) file and then list its column indices as space-separated values. Because the entire file must be read in from disk for processing, the initialization of data structures such as the row offsets can be computed at runtime. The partitioning offsets are still needed, however, and these are supplied through a separate \texttt{.dist} (distribution) file.

As additional information to a geometric partitioner, we also initialize a \texttt{.coord.k} (coordinate) file corresponding to the spatial coordinates of a vertex $(x,y,z)$ within a cartesian coordinate system. This becomes especially useful when network sizes exceed the memory requirements for advanced partitioners and may need to fall back to simple voxel-based partitioning.

For the main network \texttt{.state.k} (state) files, we supply a space-separated list of string-based model identifiers and tuples of state data. We opt to colocate the vertex and edge models together, resulting in a file that begins with a vertex identifier and its state, and followed by edge identifiers and state for each of the incoming connections. Of note here is that because the adjacency file for graph partitioning is typically undirected as opposed to directed, we additionally include special `none' model identifiers with no associated state for edge where there is an outgoing connection but no incoming one. As mentioned previously, we also introduce a \texttt{.model} file which provides a mapping between the string-based model identifiers and the size of its state tuple, as well as shared model parameters.

There are also \texttt{.event.k} (event) files which provide serialization of any simulation events ``in-flight'' that have not yet been processed on the target vertex due to connection delays. These include space-separated tuples of the source vertex, arrival time, the event type, and any associated data.

Here, we point to an implementation of this distributed file format in the Simulation Tool for Asynchronous Cortical Streams (STACS) simulator~\cite{Wang:2015}. Incidentally, STACS was built from the ground up for parallel simulation using the Charm++ parallel programming framework which essentially forces serialization for the packing-unpacking of messages between parallel objects to support fault-tolerant computing~\cite{Kale:1993}. In effect, this enabled the decoupling of the network generation process and the simulation process through an intermediate serialized representation. It also served as an efficient format to snapshot of the network state for checkpointing-restarting and offline analysis.

As a comparative scalability example, we built and serialized the cortical microcircuit model consisting of roughly 76K neurons and 0.3B synapses~\cite{Potjans:2014}, resulting in about 12GB on disk (regardless of the number of partitions). For a 2x (in neurons) for 154K neurons and 1.2B synapses, our result was about 49GB on disk. This is effectively linear cost in number of synapses.

\section{Discussion}\label{sec:discussion}
What makes the extended dCSR format particularly appealing is that it draws from pre-existing, widely used formats that are intuitive to understand. For our implementation, all of the network state is serialized into essentially four main types of parallel files (adjacency, coordinate, state, and event), and there is also little overhead in storage costs with the introduction of few additional metadata files (dist, model). Due to its simplicity, it also becomes relatively straightforward to interoperate with popular graph analysis packages such as NetworkX and its directed graph data structure~\cite{Nx:2008}.

Beyond its simplicity, we also contend that a partition-based distribution of network state makes dCSR more immediately suitable for computational parallelism, whether its target platform is between different nodes for simulation or between different chips on neuromorphic hardware. Furthermore, as a result of its lineage in graph partitioners, such a serialization may also be readily used to inform a potential repartitioning of an SNN model such that it may optimally fit to different backends. For these reasons, we put forward the extended dCSR data format for adoption within the neural computing community.

\begin{acks}
Sandia National Laboratories is a multimission laboratory managed and operated by National Technology \& Engineering Solutions of Sandia, LLC, a wholly owned subsidiary of Honeywell International Inc., for the U.S. Department of Energy’s National Nuclear Security Administration under contract DE-NA0003525.\\
This paper describes objective technical results and analysis. Any subjective views or opinions that might be expressed in the paper do not necessarily represent the views of the U.S. Department of Energy or the United States Government.
\end{acks}

\bibliographystyle{ACM-Reference-Format}
\bibliography{felwang-nice2023}


\begin{thebibliography}{24}


\ifx \showCODEN    \undefined \def \showCODEN     #1{\unskip}     \fi
\ifx \showDOI      \undefined \def \showDOI       #1{#1}\fi
\ifx \showISBNx    \undefined \def \showISBNx     #1{\unskip}     \fi
\ifx \showISBNxiii \undefined \def \showISBNxiii  #1{\unskip}     \fi
\ifx \showISSN     \undefined \def \showISSN      #1{\unskip}     \fi
\ifx \showLCCN     \undefined \def \showLCCN      #1{\unskip}     \fi
\ifx \shownote     \undefined \def \shownote      #1{#1}          \fi
\ifx \showarticletitle \undefined \def \showarticletitle #1{#1}   \fi
\ifx \showURL      \undefined \def \showURL       {\relax}        \fi
\providecommand\bibfield[2]{#2}
\providecommand\bibinfo[2]{#2}
\providecommand\natexlab[1]{#1}
\providecommand\showeprint[2][]{arXiv:#2}

\bibitem[\protect\citeauthoryear{Aimone, Severa, and Vineyard}{Aimone
  et~al\mbox{.}}{2019}]%
        {Aimone:2019}
\bibfield{author}{\bibinfo{person}{James~B. Aimone}, \bibinfo{person}{William
  Severa}, {and} \bibinfo{person}{Craig~M. Vineyard}.}
  \bibinfo{year}{2019}\natexlab{}.
\newblock \showarticletitle{Composing Neural Algorithms with Fugu}. In
  \bibinfo{booktitle}{\emph{Proceedings of the International Conference on
  Neuromorphic Systems}} (Knoxville, TN, USA) \emph{(\bibinfo{series}{ICONS
  '19})}. \bibinfo{publisher}{Association for Computing Machinery},
  \bibinfo{address}{New York, NY, USA}, Article \bibinfo{articleno}{3},
  \bibinfo{numpages}{8}~pages.
\newblock
\showISBNx{9781450376808}
\urldef\tempurl%
\url{https://doi.org/10.1145/3354265.3354268}
\showDOI{\tempurl}


\bibitem[\protect\citeauthoryear{Bai, Lu, Zhang, et~al\mbox{.}}{Bai
  et~al\mbox{.}}{2019}]%
        {Bai:2019}
\bibfield{author}{\bibinfo{person}{Junjie Bai}, \bibinfo{person}{Fang Lu},
  \bibinfo{person}{Ke Zhang}, {et~al\mbox{.}}} \bibinfo{year}{2019}\natexlab{}.
\newblock \bibinfo{title}{ONNX: Open Neural Network Exchange}.
\newblock \bibinfo{howpublished}{\url{https://github.com/onnx/onnx}}.
\newblock


\bibitem[\protect\citeauthoryear{Brette, Rudolph, Carnevale, Hines, Beeman,
  Bower, Diesmann, Morrison, Goodman, Harris, Zirpe, Natschl{\"a}ger, Pecevski,
  Ermentrout, Djurfeldt, Lansner, Rochel, Vieville, Muller, Davison,
  El~Boustani, and Destexhe}{Brette et~al\mbox{.}}{2007}]%
        {Brette:2007}
\bibfield{author}{\bibinfo{person}{Romain Brette}, \bibinfo{person}{Michelle
  Rudolph}, \bibinfo{person}{Ted Carnevale}, \bibinfo{person}{Michael Hines},
  \bibinfo{person}{David Beeman}, \bibinfo{person}{James~M Bower},
  \bibinfo{person}{Markus Diesmann}, \bibinfo{person}{Abigail Morrison},
  \bibinfo{person}{Philip~H Goodman}, \bibinfo{person}{Frederick C~Jr Harris},
  \bibinfo{person}{Milind Zirpe}, \bibinfo{person}{Thomas Natschl{\"a}ger},
  \bibinfo{person}{Dejan Pecevski}, \bibinfo{person}{Bard Ermentrout},
  \bibinfo{person}{Mikael Djurfeldt}, \bibinfo{person}{Anders Lansner},
  \bibinfo{person}{Olivier Rochel}, \bibinfo{person}{Thierry Vieville},
  \bibinfo{person}{Eilif Muller}, \bibinfo{person}{Andrew~P Davison},
  \bibinfo{person}{Sami El~Boustani}, {and} \bibinfo{person}{Alain Destexhe}.}
  \bibinfo{year}{2007}\natexlab{}.
\newblock \showarticletitle{Simulation of networks of spiking neurons: a review
  of tools and strategies.}
\newblock \bibinfo{journal}{\emph{Journal of computational neuroscience}}
  \bibinfo{volume}{23}, \bibinfo{number}{3} (\bibinfo{date}{Dec}
  \bibinfo{year}{2007}), \bibinfo{pages}{349--398}.
\newblock


\bibitem[\protect\citeauthoryear{Dai and et~al}{Dai and et~al}{2020}]%
        {Dai:2020}
\bibfield{author}{\bibinfo{person}{Kael Dai} {and} \bibinfo{person}{et al}.}
  \bibinfo{year}{2020}\natexlab{}.
\newblock \showarticletitle{The {SONATA} data format for efficient description
  of large-scale network models.}
\newblock \bibinfo{journal}{\emph{PLOS Computational Biology}}
  \bibinfo{volume}{16}, \bibinfo{number}{2} (\bibinfo{year}{2020}),
  \bibinfo{pages}{e1007696}.
\newblock


\bibitem[\protect\citeauthoryear{Dai, Gratiy, Billeh, Xu, Cai, Cain, Rimehaug,
  Stasik, Einevoll, Mihalas, Koch, and Arkhipov}{Dai et~al\mbox{.}}{2020}]%
        {Dai:2020b}
\bibfield{author}{\bibinfo{person}{Kael Dai}, \bibinfo{person}{Sergey~L.
  Gratiy}, \bibinfo{person}{Yazan~N. Billeh}, \bibinfo{person}{Richard Xu},
  \bibinfo{person}{Binghuang Cai}, \bibinfo{person}{Nicholas Cain},
  \bibinfo{person}{Atle~E. Rimehaug}, \bibinfo{person}{Alexander~J. Stasik},
  \bibinfo{person}{Gaute~T. Einevoll}, \bibinfo{person}{Stefan Mihalas},
  \bibinfo{person}{Christof Koch}, {and} \bibinfo{person}{Anton Arkhipov}.}
  \bibinfo{year}{2020}\natexlab{}.
\newblock \showarticletitle{Brain Modeling ToolKit: An open source software
  suite for multiscale modeling of brain circuits}.
\newblock \bibinfo{journal}{\emph{PLoS Computational Biology}}
  \bibinfo{volume}{16} (\bibinfo{year}{2020}).
\newblock


\bibitem[\protect\citeauthoryear{Davies}{Davies}{2019}]%
        {Davies:2019}
\bibfield{author}{\bibinfo{person}{Mike Davies}.}
  \bibinfo{year}{2019}\natexlab{}.
\newblock \showarticletitle{Benchmarks for progress in neuromorphic computing}.
\newblock \bibinfo{journal}{\emph{Nature Machine Intelligence}}
  \bibinfo{volume}{1}, \bibinfo{number}{9} (\bibinfo{year}{2019}),
  \bibinfo{pages}{386--388}.
\newblock


\bibitem[\protect\citeauthoryear{Davison, Br{\"u}derle, Eppler, Kremkow,
  Muller, Pecevski, Perrinet, and Yger}{Davison et~al\mbox{.}}{2009}]%
        {Davison:2009}
\bibfield{author}{\bibinfo{person}{Andrew Davison}, \bibinfo{person}{Daniel
  Br{\"u}derle}, \bibinfo{person}{Jochen Eppler}, \bibinfo{person}{Jens
  Kremkow}, \bibinfo{person}{Eilif Muller}, \bibinfo{person}{Dejan Pecevski},
  \bibinfo{person}{Laurent Perrinet}, {and} \bibinfo{person}{Pierre Yger}.}
  \bibinfo{year}{2009}\natexlab{}.
\newblock \showarticletitle{PyNN: a common interface for neuronal network
  simulators}.
\newblock \bibinfo{journal}{\emph{Frontiers in Neuroinformatics}}
  \bibinfo{volume}{2} (\bibinfo{year}{2009}).
\newblock
\showISSN{1662-5196}
\urldef\tempurl%
\url{https://doi.org/10.3389/neuro.11.011.2008}
\showDOI{\tempurl}


\bibitem[\protect\citeauthoryear{Golosio, Tiddia, De~Luca, Pastorelli, Simula,
  and Paolucci}{Golosio et~al\mbox{.}}{2021}]%
        {Golosio:2021}
\bibfield{author}{\bibinfo{person}{Bruno Golosio}, \bibinfo{person}{Gianmarco
  Tiddia}, \bibinfo{person}{Chiara De~Luca}, \bibinfo{person}{Elena
  Pastorelli}, \bibinfo{person}{Francesco Simula}, {and}
  \bibinfo{person}{Pier~Stanislao Paolucci}.} \bibinfo{year}{2021}\natexlab{}.
\newblock \showarticletitle{Fast Simulations of Highly-Connected Spiking
  Cortical Models Using GPUs}.
\newblock \bibinfo{journal}{\emph{Frontiers in Computational Neuroscience}}
  \bibinfo{volume}{15} (\bibinfo{year}{2021}).
\newblock


\bibitem[\protect\citeauthoryear{Hagberg, Schult, and Swart}{Hagberg
  et~al\mbox{.}}{2008}]%
        {Nx:2008}
\bibfield{author}{\bibinfo{person}{Aric~A. Hagberg}, \bibinfo{person}{Daniel~A.
  Schult}, {and} \bibinfo{person}{Pieter~J. Swart}.}
  \bibinfo{year}{2008}\natexlab{}.
\newblock \showarticletitle{Exploring Network Structure, Dynamics, and Function
  using NetworkX}. In \bibinfo{booktitle}{\emph{Proceedings of the 7th Python
  in Science Conference}}, \bibfield{editor}{\bibinfo{person}{Ga\"el
  Varoquaux}, \bibinfo{person}{Travis Vaught}, {and} \bibinfo{person}{Jarrod
  Millman}} (Eds.). \bibinfo{address}{Pasadena, CA USA}, \bibinfo{pages}{11 --
  15}.
\newblock


\bibitem[\protect\citeauthoryear{Intel}{Intel}{2022}]%
        {Lava:2022}
\bibfield{author}{\bibinfo{person}{Intel}.} \bibinfo{year}{2022}\natexlab{}.
\newblock \bibinfo{title}{Lava: A Software Framework for Neuromorphic
  Computing}.
\newblock \bibinfo{howpublished}{GitHub Repository}.
\newblock
\urldef\tempurl%
\url{https://github.com/lava-nc/lava}
\showURL{%
\tempurl}


\bibitem[\protect\citeauthoryear{Kale and Krishnan}{Kale and Krishnan}{1993}]%
        {Kale:1993}
\bibfield{author}{\bibinfo{person}{Laxmikant~V. Kale} {and}
  \bibinfo{person}{Sanjeev Krishnan}.} \bibinfo{year}{1993}\natexlab{}.
\newblock \showarticletitle{CHARM++: A Portable Concurrent Object Oriented
  System Based on C++}. In \bibinfo{booktitle}{\emph{Proceedings of the Eighth
  Annual Conference on Object-Oriented Programming Systems, Languages, and
  Applications}} (Washington, D.C., USA) \emph{(\bibinfo{series}{OOPSLA '93})}.
  \bibinfo{publisher}{Association for Computing Machinery},
  \bibinfo{address}{New York, NY, USA}, \bibinfo{pages}{91--108}.
\newblock
\showISBNx{0897915879}
\urldef\tempurl%
\url{https://doi.org/10.1145/165854.165874}
\showDOI{\tempurl}


\bibitem[\protect\citeauthoryear{Karypis and Kumar}{Karypis and Kumar}{1998}]%
        {Karypis:1998}
\bibfield{author}{\bibinfo{person}{George Karypis} {and} \bibinfo{person}{Vipin
  Kumar}.} \bibinfo{year}{1998}\natexlab{}.
\newblock \showarticletitle{A Parallel Algorithm for Multilevel Graph
  Partitioning and Sparse Matrix Ordering}.
\newblock \bibinfo{journal}{\emph{J. Parallel and Distrib. Comput.}}
  \bibinfo{volume}{48}, \bibinfo{number}{1} (\bibinfo{year}{1998}),
  \bibinfo{pages}{71--95}.
\newblock
\showISSN{0743-7315}
\urldef\tempurl%
\url{https://doi.org/10.1006/jpdc.1997.1403}
\showDOI{\tempurl}


\bibitem[\protect\citeauthoryear{Karypis and Schloegel}{Karypis and
  Schloegel}{2013}]%
        {Parmetis:2013}
\bibfield{author}{\bibinfo{person}{George Karypis} {and} \bibinfo{person}{Kirk
  Schloegel}.} \bibinfo{year}{2013}\natexlab{}.
\newblock \bibinfo{booktitle}{\emph{ParMETIS: Parallel Graph Partitioning and
  Sparse Matrix Ordering Library} (\bibinfo{edition}{4.0} ed.)}.
\newblock University of Minnesota, Department of Computer Science and
  Engineering, Minneapolis, MN.
\newblock


\bibitem[\protect\citeauthoryear{Knight and Nowotny}{Knight and
  Nowotny}{2021}]%
        {Knight:2021}
\bibfield{author}{\bibinfo{person}{James~C. Knight} {and}
  \bibinfo{person}{Thomas Nowotny}.} \bibinfo{year}{2021}\natexlab{}.
\newblock \showarticletitle{Larger GPU-accelerated brain simulations with
  procedural connectivity}.
\newblock \bibinfo{journal}{\emph{Nature Computational Science}}
  \bibinfo{volume}{1}, \bibinfo{number}{2} (\bibinfo{year}{2021}),
  \bibinfo{pages}{136--142}.
\newblock
\showISBNx{2662-8457}
\urldef\tempurl%
\url{https://doi.org/10.1038/s43588-020-00022-7}
\showDOI{\tempurl}


\bibitem[\protect\citeauthoryear{Kulkarni, Parsa, Mitchell, and
  Schuman}{Kulkarni et~al\mbox{.}}{2021}]%
        {Kulkarni:2021}
\bibfield{author}{\bibinfo{person}{S. Kulkarni}, \bibinfo{person}{M. Parsa},
  \bibinfo{person}{J.~P. Mitchell}, {and} \bibinfo{person}{C.~D. Schuman}.}
  \bibinfo{year}{2021}\natexlab{}.
\newblock \showarticletitle{Benchmarking the Performance of Neuromorphic and
  Spiking Neural Simulators}.
\newblock \bibinfo{journal}{\emph{Neurocomputing}}  \bibinfo{volume}{447}
  (\bibinfo{year}{2021}), \bibinfo{pages}{145--160}.
\newblock


\bibitem[\protect\citeauthoryear{Pehle and Pedersen}{Pehle and
  Pedersen}{2021}]%
        {Norse:2021}
\bibfield{author}{\bibinfo{person}{Christian Pehle} {and}
  \bibinfo{person}{Jens~Egholm Pedersen}.} \bibinfo{year}{2021}\natexlab{}.
\newblock \bibinfo{title}{{Norse - A deep learning library for spiking neural
  networks}}.
\newblock
\newblock
\urldef\tempurl%
\url{https://doi.org/10.5281/zenodo.4422025}
\showDOI{\tempurl}
\newblock
\shownote{Documentation: https://norse.ai/docs/.}


\bibitem[\protect\citeauthoryear{Potjans and Diesmann}{Potjans and
  Diesmann}{2014}]%
        {Potjans:2014}
\bibfield{author}{\bibinfo{person}{Tobias~C Potjans} {and}
  \bibinfo{person}{Markus Diesmann}.} \bibinfo{year}{2014}\natexlab{}.
\newblock \showarticletitle{The cell-type specific cortical microcircuit:
  relating structure and activity in a full-scale spiking network model.}
\newblock \bibinfo{journal}{\emph{Cereb Cortex}} \bibinfo{volume}{24},
  \bibinfo{number}{3} (\bibinfo{date}{Mar} \bibinfo{year}{2014}),
  \bibinfo{pages}{785--806}.
\newblock


\bibitem[\protect\citeauthoryear{Rothganger, Warrender, Trumbo, and
  Aimone}{Rothganger et~al\mbox{.}}{2014}]%
        {Rothganger:2014}
\bibfield{author}{\bibinfo{person}{Fredrick Rothganger},
  \bibinfo{person}{Christina Warrender}, \bibinfo{person}{Derek Trumbo}, {and}
  \bibinfo{person}{James Aimone}.} \bibinfo{year}{2014}\natexlab{}.
\newblock \showarticletitle{N2A: a computational tool for modeling from neurons
  to algorithms}.
\newblock \bibinfo{journal}{\emph{Frontiers in Neural Circuits}}
  \bibinfo{volume}{8} (\bibinfo{year}{2014}).
\newblock
\showISSN{1662-5110}
\urldef\tempurl%
\url{https://doi.org/10.3389/fncir.2014.00001}
\showDOI{\tempurl}


\bibitem[\protect\citeauthoryear{Saad}{Saad}{2011}]%
        {Saad:2011}
\bibfield{author}{\bibinfo{person}{Yousef Saad}.}
  \bibinfo{year}{2011}\natexlab{}.
\newblock \bibinfo{booktitle}{\emph{Numerical Methods for Large Eigenvalue
  Problems}}.
\newblock \bibinfo{publisher}{Society for Industrial and Applied Mathematics}.
\newblock
\urldef\tempurl%
\url{https://doi.org/10.1137/1.9781611970739}
\showDOI{\tempurl}
\showeprint{https://epubs.siam.org/doi/pdf/10.1137/1.9781611970739}


\bibitem[\protect\citeauthoryear{Schuman, Potok, Patton, Birdwell, Dean, Rose,
  and Plank}{Schuman et~al\mbox{.}}{2017}]%
        {Schuman:2017}
\bibfield{author}{\bibinfo{person}{Catherine~D. Schuman},
  \bibinfo{person}{Thomas~E. Potok}, \bibinfo{person}{Robert~M. Patton},
  \bibinfo{person}{J.~Douglas Birdwell}, \bibinfo{person}{Mark~E. Dean},
  \bibinfo{person}{Garrett~S. Rose}, {and} \bibinfo{person}{James~S. Plank}.}
  \bibinfo{year}{2017}\natexlab{}.
\newblock \showarticletitle{A Survey of Neuromorphic Computing and Neural
  Networks in Hardware}.
\newblock \bibinfo{journal}{\emph{ArXiv}}  \bibinfo{volume}{abs/1705.06963}
  (\bibinfo{year}{2017}).
\newblock


\bibitem[\protect\citeauthoryear{Stimberg, Brette, and Goodman}{Stimberg
  et~al\mbox{.}}{2019}]%
        {Stimberg:2019}
\bibfield{author}{\bibinfo{person}{Marcel Stimberg}, \bibinfo{person}{Romain
  Brette}, {and} \bibinfo{person}{Dan~FM Goodman}.}
  \bibinfo{year}{2019}\natexlab{}.
\newblock \showarticletitle{Brian 2, an intuitive and efficient neural
  simulator}.
\newblock \bibinfo{journal}{\emph{eLife}}  \bibinfo{volume}{8}
  (\bibinfo{date}{Aug.} \bibinfo{year}{2019}), \bibinfo{pages}{e47314}.
\newblock
\showISSN{2050-084X}
\urldef\tempurl%
\url{https://doi.org/10.7554/eLife.47314}
\showDOI{\tempurl}


\bibitem[\protect\citeauthoryear{Vineyard, Green, Severa, and Ko\c{c}}{Vineyard
  et~al\mbox{.}}{2019}]%
        {Vineyard:2019}
\bibfield{author}{\bibinfo{person}{Craig~M. Vineyard}, \bibinfo{person}{Sam
  Green}, \bibinfo{person}{William~M. Severa}, {and}
  \bibinfo{person}{\c{C}etin~Kaya Ko\c{c}}.} \bibinfo{year}{2019}\natexlab{}.
\newblock \showarticletitle{Benchmarking Event-Driven Neuromorphic
  Architectures}. In \bibinfo{booktitle}{\emph{Proceedings of the International
  Conference on Neuromorphic Systems}} (Knoxville, TN, USA).
  \bibinfo{publisher}{Association for Computing Machinery},
  \bibinfo{address}{New York, NY, USA}, \bibinfo{numpages}{5}~pages.
\newblock


\bibitem[\protect\citeauthoryear{Wang}{Wang}{2015}]%
        {Wang:2015}
\bibfield{author}{\bibinfo{person}{Felix Wang}.}
  \bibinfo{year}{2015}\natexlab{}.
\newblock \showarticletitle{Simulation Tool for Asynchronous Cortical Streams
  (STACS): Interfacing with Spiking Neural Networks}. In
  \bibinfo{booktitle}{\emph{Procedia Computer Science}} (San Jose, CA, USA),
  Vol.~\bibinfo{volume}{61}. \bibinfo{pages}{322--7}.
\newblock


\bibitem[\protect\citeauthoryear{West}{West}{2001}]%
        {West:2001}
\bibfield{author}{\bibinfo{person}{Douglas~B. West}.}
  \bibinfo{year}{2001}\natexlab{}.
\newblock \bibinfo{booktitle}{\emph{Introduction to Graph Theory}
  (\bibinfo{edition}{2} ed.)}.
\newblock \bibinfo{publisher}{Prentice Hall}.
\newblock
\showISBNx{0130144002}


\end{thebibliography}

\end{document}